\documentclass[%
 reprint,
superscriptaddress,
 amsmath,amssymb,
 aps,
prper,
]{revtex4-2}

\usepackage{dcolumn}
\usepackage{bm}
\usepackage{hyperref}
\usepackage{graphicx}
\usepackage{float}
\usepackage{amsmath}
\usepackage{amssymb}
\usepackage{array}
\usepackage{multirow}
\usepackage{longtable}
\usepackage{textgreek}
\usepackage{booktabs}
\usepackage{framed}
\usepackage{ragged2e}
\usepackage{afterpage}
\usepackage[inline]{enumitem}

\begin{document}

\preprint{APS/123-QED}

\title{Assessing physics quantitative literacy development in algebra-based physics}

\author{Charlotte Zimmerman}

\affiliation{Department of Physics, University of Washington, Box 351560, Seattle, WA 98195-1560, USA}

\author{Alexis Olsho}

\affiliation{Department of Physics and Meteorology, United States Air Force Academy, 2354 Fairchild Drive, USAF Academy, CO 80840 USA}
\author{Trevor I. Smith}

\affiliation{Department of Physics \& Astronomy and Department of STEAM Education, Rowan University, 201 Mullica Hill Rd., Glassboro, NJ 08028, USA}
\author{Philip Eaton}

\affiliation{School of Natural Sciences and Mathematics, Stockton University, Galloway, NJ 08205, USA}
\author{Suzanne White Brahmia}

\affiliation{Department of Physics, University of Washington, Box 351560, Seattle, WA 98195-1560, USA}

\date{\today}

\begin{abstract}

Quantitative reasoning is an essential learning objective of physics instruction. The Physics Inventory for Quantitative Literacy (PIQL) is an assessment tool that has been developed for calculus-based physics courses to help instructors evaluate whether their students learn to reason this way \cite{PIQLPaper}. However, the PIQL is not appropriate for the large population of students taking physics who are not enrolled in, or have not completed, calculus. To address this need, we have developed the General Equation-based Reasoning inventory of QuaNtity (GERQN). The GERQN is an \textit{algebra-based version} of the PIQL and is appropriate for most physics students; the only requirement is that students have taken algebra so they are familiar with the use of variable, negative quantities, and linear functions.  In this paper we present the development and validation of the GERQN, and a short discussion on how the GERQN can be used by instructors to help their students learn.

\end{abstract}

\maketitle

\section{Introduction}
\label{sec:intro}

Reasoning quantitatively is a learning objective of many physics courses, and reasoning about mathematical models and their physical meaning is at the heart of what physicists do. Mathematical modeling of the physical world involves generating, translating, and interpreting the physical meaning of mathematical representations, and developing the skills to do so is a valued learning outcome of an introductory physics course \cite{Uhden_Karam_Pietrocola_Pospiech_2012, 
White_2019, Gifford_Finkelstein_2020, Czocher_Hardison_2021}. 

In this work, we focus on one piece of modeling: quantitative literacy (QL). The idea of quantitative literacy was introduced by mathematics education researchers, and is related to the practice of using familiar mathematics to represent the world \cite{Thompson_2011}. 
The Physics Inventory of Quantitative Literacy (PIQL, pronounced ``pickle'') assesses \textit{physics} quantitative literacy (PQL), or quantitative literacy situated in physical contexts. It is a reasoning inventory that can be used to measure the degree to which students learn to reason quantitatively as a result of taking physics courses \cite{PIQLPaper}.  The PIQL is written for a student population that has either completed or is co-enrolled in calculus, and therefore includes reasoning about vectors, limits, and changing rates of change. Research has shown that reasoning this way is challenging for many; students do not readily saturate the PIQL even after a year of college-level instruction, and scores are unlikely to change without direct instruction \cite{PIQLPaper}.

PQL relies largely on algebraic fundamentals typically taught in middle school and early high school, such as reasoning about ratio and rates of change. There is an opportunity for this kind of reasoning to be a learning objective of earlier physics courses. However, the PIQL is already difficult for students enrolled in college-level calculus-based physics\cite{PIQLPaper}. A version of the PIQL that is specifically designed for algebra-prerequiste courses is needed. Such an instrument could then be used to help support the large number of students  enrolled in algebra-based physics at both the precollege and college level develop PQL.

To address this need, we have developed an algebra-based version of the PIQL: the Generalized Reasoning-based inventory of QuaNtity (GERQN, pronounced ``gherkin''). We intend the GERQN to expand the target population to include students who have completed algebra I; this necessarily also entails expanding the population of experts whom we expect to engage with the inventory. In this paper, we present the GERQN and describe its development and validation. This process included interviews with mathematics learning experts, physics teaching professionals in high schools, and physics experts who teach at the college level. We conclude with a short reflection on curricular implications and how the assessment could be used by instructors to inform activity development.

\section{Background}
\label{sec:background}

The GERQN relies on the same theoretical framework as the PIQL \cite{PIQLPaper}, adjusted to an algebra-based level. In this section, we first summarize the facets of PQL that were used in the development of the PIQL for calculus-based courses, and aspects that are also prevalent in algebra-based courses. We then discuss our framework for test development, and theoretical basis for the statistical measures used to establish test validity.

\subsection{Test Construct: Three Facets of PQL}
We define quantitative literacy as ``the interconnected skills, attitudes, and habits of mind that together support the sophisticated use of familiar mathematics for sense-making'' \cite{PIQLPaper,Thompson_2011,Ojose_2011}. \textit{Physics} quantitative literacy (PQL) refers to quantitative literacy situated in physics contexts. In this way, PQL represents a conceptual blend between physics and mathematical reasoning \cite{Bing_Redish_2007, Hu_Rebello_2013, VandenEynde_Schermerhorn_Deprez_Goedhart_Thompson_DeCock_2020}. The PIQL---the Physics Inventory of Quantitative Literacy---was developed to address the need for a validated reasoning inventory that can help instructors measure student development of PQL \cite{PIQLPaper}. It was designed to measure three main facets of PQL: reasoning about signs, covariational reasoning, and proportional reasoning.

Reasoning in physics about the meaning of sign and signed quantities is nuanced \cite{WhiteBrahmia_Olsho_Smith_Boudreaux_2020}. For example, a negative sign could mean an amount of a quantity with respect to zero, a decrease in amount, a direction, or a type of charge -- just to name a few. Mathematics education research has attended to the challenges students face when interpreting the negative sign; physics education research has expanded on this work to describe the additional difficulty of interpreting the meaning of negative physical quantities \cite{Vlassis_2008, Bajracharya_Sealey_Thompson_2023, Ceuppens_Bollen_Deprez_Dehaene_DeCock_2019, olsho2021negative}. PQL associated with signed quantities includes recognizing the physical interpretation of the negative sign in various contexts, and distinguishing between negative quantities and negative rates of change. It also includes flexibly incorporating negative signs when translating between symbolic, graphical, and written representations of physical scenarios.

The use of vector quantities, and attention to their sign symbolically, varies considerably from algebra-based to calculus-based physics courses. For example, while dot products and cross products are common in calculus-based courses, algebra-based courses typically only include vector addition and decomposition \cite{APPhyscs1}. Treatment of Hooke's law provides another common example---while calculus-based textbooks are likely to include the negative sign in the definition ($\vec{F} = k\Delta \vec{x}$), algebra-based textbooks often treat the discussion of the negative sign as conceptual and provide a definition based on magnitudes ($F = k\Delta x$) \cite{Mazur_2015, Knight_Jones_Field_2019}.

Covariational reasoning refers to reasoning about the change in one quantity with respect to the change in another, related quantity. It is commonly used in research on undergraduate mathematics education when studying student reasoning in pre-calculus and calculus courses \cite{Carlson_Jacobs_Coe_Larsen_Hsu_2002,Thompson_Carlson_2017, Jones_2022}. In the physics education research literature,  covariational reasoning is most often referred to as ``scaling'' \cite{Arons_1976, Trowbridge_McDermott_1981, Bissell_Ali_Postle_2022}. However, recent work has leveraged the language of covariation to consider more specifically how students and experts describe changing quantities in physics and across STEM more broadly \cite{Zimmerman_Olsho_Loverude_WhiteBrahmia_2024, Emigh_Siegel_Alfson_Gire_2019, VandenEynde_vanKampen_VanDooren_DeCock_2019, Altindis_Bowe_Couch_Bauer_Aikens_2024}. 

Mathematics education researchers have developed and iterated on a framework for covariation that clearly identifies specific types of covariation and activities associated with each type \cite{Carlson_Jacobs_Coe_Larsen_Hsu_2002, Thompson_Carlson_2017, Jones_2022}. Recent work in physics education has built on this framework to operationalize covariational reasoning specifically with physics quantities \cite{Olsho_Zimmerman_Boudreaux_Smith_Eaton_WhiteBrahmia_2022}. The PIQL contains covariational reasoning items that ask students to consider changing rates of change symbolically and graphically, reasoning about discrete covariation with multiple variables (``If this quantity doubles and another is tripled, what happens to a third quantity?''), and reasoning using limits.

Proportional reasoning has been studied in physics education research as ratio or as describing a function-based relationship between two quantities \cite{Akatugba_Wallace_1999, Maloney_Hieggelke_Kanim_2010, Sherin_2001}. Research has described several specific ways that proportional reasoning is foundational to introductory physics, including proportion as unit rate (i.e. this amount of one quantity ``for every'' unit of that quantity) and using a ratio to determine an unknown quantity \cite{Boudreaux_Kanim_Olsho_WhiteBrahmia_Zimmerman_Smith_2020}. Historically, proportional reasoning has been used in physics education research to describe many different kinds of function-based relationships between quantities. In this work, we consider proportional reasoning to be the linear subset of covariational reasoning. That is, proportional reasoning describes covariational reasoning about quantities related by a linear function. 

We include proportional reasoning as a facet of the same importance as reasoning about sign and covariational reasoning because of the high prevalence of linear relationships in introductory physics. This is especially true in algebra-based introductory physics, where more complex functions are often left for future courses. For example, it is common to treat drag forces that are non-linear with respect to velocity only in calculus-based courses. It is also common practice to discuss changing rates of change qualitatively in algebra-based courses, but not expect students enrolled in such courses to solve symbolic expressions of non-linear functions for their rate of change. For example, non-constant acceleration is rarely taught using symbolic representations and procedures in an algebra-based course \cite{Mazur_2015, Knight_Jones_Field_2019}. However, interpreting the meaning of graphical representations of changing rates of change is often considered within the scope.

Our aim in developing the GERQN was to create an assessment to measure reasoning about all three facets of PQL, but at an appropriate level for students who have completed algebra I. Research about pre-calculus and calculus resources has argued that much of the reasoning we associate with calculus stems from reasoning developed in algebra---namely, the meaning of symbols and reasoning about changes in quantities (covariational reasoning), with a particular focus on constant rates of change (proportional reasoning) \cite{Redish_Kuo_2015,Uhden_Karam_Pietrocola_Pospiech_2012,Boudreaux_Kanim_WhiteBrahmia_2015,Olsho_WhiteBrahmia_Boudreaux_Smith_2019,Sherin_2001,VandenEynde_vanKampen_VanDooren_DeCock_2019,WhiteBrahmia_2014,WhiteBrahmia_Olsho_Boudreaux_Smith_Zimmerman_2020}. Therefore, we consider our aim to be to measure \textit{calculus-like} ideas without requiring the symbolic infrastructure, procedures, or reasoning about infinitesimal change that one may associate with a calculus-based course.

\subsection{Test Development}
We used the protocol described by Adams and Wieman \cite{Adams_Wieman_2011} to guide the development and validation process of the GERQN. The Adams and Wieman framework is intended for creation of a ``formative assessment of instruction''---that is, an instrument that can be used to measure changes in student reasoning as a result of instruction \textit{in order to inform instructional change}. The mixed-methods approach proposed by Adams and Wieman is commonly followed (see, for example, references \cite{zhang2024developing, Santoso_Istiyono_Haryanto_Retnawati_2024, Holt_Duke_Dunk_and_2024, McKee_Orlov_2023, Cetnar_Besemer_Bry_Buckey_Burmeister_Rodrigues_Schubert_Speidel_Sutlief_Yu_2023}), and is particularly well-suited for guiding the development of instruments such as the GERQN that involves student and expert reasoning currently not well-represented in the physics education research literature. The approach includes multiple rounds of interviews with students, instructors, and other content experts, combined with statistical analyses of individual assessment items and the instrument as a whole. The framework describes a clear process for development of a valid and reliable instrument that can provide a measure of changes in students' PQL over the course of instruction in introductory physics. Moreover, collection of qualitative data during the validation process can ensure that the instrument will be valued by instructors, which can lead to strong uptake and therefore improvement in educational outcomes. 

Statistical measures from classical test theory (CTT) can provide quantitative evidence for the validity and reliability of the GERQN \cite{PIQLPaper,Adams_Wieman_2011,Wiersma1990,Doran1980,Ding2009}. Assembling a test containing items with a broad range of item difficulties and high values of item discrimination (or the point-biserial coefficient) provides evidence of validity that the test can accurately measure a broad range of students reasoning. High values of test-level statistics such as Cronbach's $\alpha$ and Ferguson's $\delta$ provide evidence of the reliability of the test, as well as justification for interpreting a single-value test score as a meaningful representation of student reasoning. In addition to CTT statistics, exploratory and confirmatory factor analyses provide evidence for whether the test measures a single knowledge/reasoning construct or several distinct constructs \cite{PIQLPaper,Brown2014,Hayton2004}.

\section{Methods}
\label{sec:methods}

\begin{figure*}[tb]
    \centering
    \includegraphics[width=\linewidth]{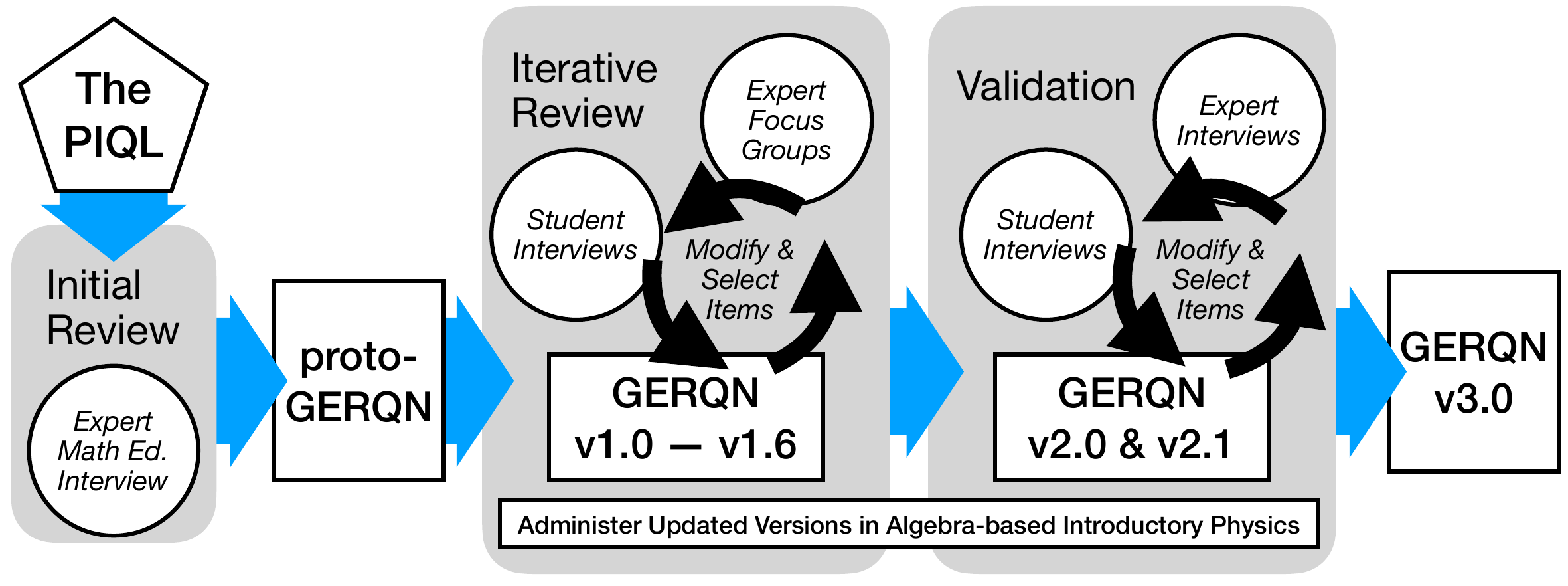}
    \caption{The development and validation process of the GERQN.}
    \label{fig:timeline}
\end{figure*}

We designed the GERQN by modifying PIQL items, developing new items as needed, and removing items that were beyond the scope of the intended audience. We followed the procedure described by Adams and Wieman and used in the development of the PIQL  \cite{Adams_Wieman_2011,PIQLPaper}. Figure~\ref{fig:timeline} describes the three stages of development and validation:
\begin{enumerate}
    \item Initial review: we examined and modified the PIQL through the lens of algebra-based reasoning, based on prior research and an interview with a middle school mathematics education researcher. The result was a prototype, or ``protoGERQN.''
    \item Iterative review: we modified, removed, and developed new items based on student interviews, an expert focus group, and item statistics. This process resulted in versions 1.0-1.6. 
    \item Validation: we conducted student validation interviews, expert validation interviews, and a final review of the item and inventory statistics. This process resulted in versions 2.0-3.0.
\end{enumerate}

The majority of our data were collected at a large, R1 university in the Pacific Northwest, which we will refer to as the main institution. Three courses make up the year-long, introductory, algebra-based physics sequence at that institution. We refer to these courses as ``Mechanics,'' ``Electricity and Magnetism,'' and ``Waves and Optics.'' We identify student data based on these course names. For example, student interviews are labeled by the course students were either enrolled in or most recently completed at the time of the interview. We collected quantitative data at the start of each term; thus, we identify the administration data sets as ``PreMech'' (which corresponds to data collected in the first two weeks of instruction in Mechanics), ``PostMech'' (which corresponds to data collected in the first two weeks of instruction in Electricity and Magnetism), and ``PostEM'' (which corresponds to data collected in the first two weeks of instruction in Waves and Optics). A summary of the data collected in each stage can be found in Table~\ref{tab:dataSummary}.

\begin{table}
    \centering
    \vspace{\baselineskip}
    \begin{tabular}{l|l c r}
        & Data Type                                 & Version & N \\
    \hline
    \hline
    \parbox[t]{3mm}{\multirow{18}{*}{\rotatebox[origin=c]{90}{Iterative Review}}}
        & \textbf{Test Admin.}               & \textbf{1.1A/B} & \textbf{708}\\
            & \hspace{1em} PreMech                  && 488 \\
            & \hspace{1em} PostMech                 && 157 \\
            & \hspace{1em} PostEM                   && 63 \\
        & \textbf{Faculty Focus Group}       & \textbf{1.1A} & \textbf{1 grp}\\
            & \hspace{1em} R1 Research (Physics)    && 2 \\
            & \hspace{1em} R1 Teaching (Physics)    && 4 \\
        & \textbf{Test Admin.}               & \textbf{1.4A/B} & \textbf{726}\\
            & \hspace{1em} PreMech                  && 231 \\
            & \hspace{1em} PostMech                 && 425 \\
            & \hspace{1em} PostEM                   && 70 \\
        & \textbf{Student Interviews}        & \textbf{1.5} & \textbf{12} \\
            & \hspace{1em} PreMech                  && 7 \\
            & \hspace{1em} PostMech                 && 5 \\
        & \textbf{Test Admin.}               & \textbf{1.6} & \textbf{2778}\\
            & \hspace{1em} PreMech                  && 1116 \\
            & \hspace{1em} PostMech                 && 864\\
            & \hspace{1em} PostEM                   && 798\\
    \hline
    \parbox[t]{3mm}{\multirow{13}{*}{\rotatebox[origin=c]{90}{Validation}}}
        & \textbf{Student Interviews}       & \textbf{1.6} & \textbf{17} \\
            & \hspace{1em} PreMech                  && 6\\
            & \hspace{1em} PostMech                 && 5\\
            & \hspace{1em} PostEM                   && 6\\
        & \textbf{Faculty Interviews}       & \textbf{2.0-2.1} & \textbf{6}\\
            & \hspace{1em} High school (Physics)      && 3\\
            & \hspace{1em} Two-year College (Physics) && 1\\
            & \hspace{1em} HBCU Research  (Physics)   && 1\\
            & \hspace{1em} R1 Research (Math Ed.)     && 1\\
        & \textbf{Test Admin.}               & \textbf{3.0} & \textbf{1453} \\
            & \hspace{1em} PreMech                  && 746 \\
            & \hspace{1em} PostMech                 && 573 \\
            & \hspace{1em} PostEM                   && 134 \\
    \hline
    \end{tabular}
    \caption{Data sources and N-values for each type of data collected during iterative review and validation. Versions 1.1 and 1.4 were each further split into two versions, A and B, equally and randomly across the student populations listed to test early drafts of item revisions.}
    \label{tab:dataSummary}
\end{table}

\subsection{Student Interviews}
We conducted two rounds of individual interviews with students enrolled in the introductory algebra-based physics sequence at the main institution. Round 1 interviews were analyzed during iterative review, and round 2 interviews were used for validation purposes. We used the same protocol for both rounds. Student interviews gave insight into whether students chose the correct answers for the correct reasons, whether common answer options were missing, and what reasoning students used. 

We recruited students via announcements within course-management software and selected participants on a first-come-first-served basis. We continued to recruit students until we had a representative sample across each course in the sequence and across demographic lines including gender, race and ethnicity, and prior experience in math and physics courses. Students were offered \$15 gift cards for participating. 

Student interviews were semi-structured and consisted of one participant and one interviewer. The participant was asked to complete the GERQN while thinking out loud, and the interviewer only spoke to prompt them to remember to talk out loud if they fell silent. After the participant indicated they were finished, the interviewer asked follow up questions to clarify the participant's reasoning. Interviews were audio-recorded and transcribed using Otter.ai software \cite{otter}. These transcripts were subsequently hand-corrected. Any written work produced by students during the interview was also collected.

Interviews were analyzed by a subset of the research team. For each item, a subset of the research team coded responses based on whether the item was answered correctly, and the reasoning used by the student to arrive at their answer. These codes were confirmed by a separate research team member who coded a subset of the data individually, and discussed with the initial coders until consensus was reached. The results were then discussed across the research team to determine whether the question prompt and/or answer choices needed to be revised, or whether to remove the item completely. Responses by students who had not taken calculus were more heavily considered in evaluating the mathematics level of the items.

\subsection{Expert Interviews}
We conducted two rounds of expert interviews: one focus group during iterative review, and a set of individual interviews during validation. For the focus group, faculty who had experience teaching in the algebra-based sequence at the main institution were recruited. We included both research-focused and teaching-focused faculty. Members of the research team reached out to candidates directly; all of those who were interested and willing were invited to participate. Participants were asked to complete the GERQN on their own before the interview, and then met together at one time with two members of the research team. The faculty focus group was intended to provide insight into whether the inventory as a whole was considered valuable by instructors, and whether it would be a useful measure of their students' reasoning. We also asked instructors for feedback on the scope of individual items for the student population they teach. 

In the validation phase, we conducted individual faculty interviews with experienced instructors across a variety of institutions. Interview candidates were identified by their experience teaching, implementation of research-validated methods, and experience with teacher preparation. Our aim was to populate the expert pool with the breadth of instructors we expect to use the GERQN. Therefore, we included several high school physics teachers (in private and public institutions), a community college physics faculty member, a physics faculty member situated at a historically black university in the American South, and a mathematics education researcher who specializes in teacher preparation around algebra and calculus reasoning. None of the experts who were interviewed for validation had been previously interviewed about the GERQN. We did not re-interview faculty at the main institution individually, as they had already been interviewed and been administering updated versions of the GERQN in their courses---we consider this evidence they already perceived the GERQN as valuable.

Faculty were asked to read the GERQN before their interview. The interviews consisted of 1--3 members of the research team and one faculty member, and were conducted and recorded over Zoom. Participants were asked if the test aligned with learning outcomes of their courses, whether there were any individual items that were too difficult or did not align with their goals of the course, and for feedback with respect to readability.  Most of the experts we interviewed also provided, unprompted, item-by-item feedback on how their students would interact with the item and suggestions for possible improvements.

\subsection{Test Administration}
The test was administered online using Qualtrics, as part of normal course activities. Students were expected to complete the assessment for a small amount of course credit based on participation. Each item was displayed on its own page, and students could move forward and backward freely with no time constraints. As revisions were made to the GERQN during the iterative review and validation stages, the most up-to-date version was administered to students each term. When there was dispute about a particular change among the research team, two versions of the GERQN were administered such that half the students in a given course saw one version, and the other half saw the other.

During the iterative review stage we examined the frequency with which students selected each response option to each item. We often removed or modified rarely selected response options, but kept response options that showed up prominently in either student or expert interviews. Additionally, we performed a variety of statistical and psychometric tests of reliability and validity using dichotomously-scored data. These included calculating item-level statistics (classical test theory difficulty and item discrimination, using the point biserial correlation), calculating test-level statistics (Cronbach's $\alpha$ and Ferguson's $\delta$), and performing both exploratory and confirmatory factor analysis.

\section{Results from Initial and Iterative Review}
\label{sec:dev}

In this section, we describe the outcomes from initial and iterative review, during the development of the GERQN.

\subsection{Initial Review}
Compared to students enrolled in calculus-based physics, the target population for the GERQN is on average at an earlier stage of their academic career and has taken fewer prior physics and mathematics courses. To determine an appropriate mathematics and reading level for the GERQN, we interviewed a middle school mathematics education researcher. She was asked to read the original PIQL and identify items containing content beyond the scope of algebra I. The discussion between the expert and the research team members resulted in removing items with reference to physics content knowledge or vectors; adjusting items with non-linear relationships to linear contexts, single variable contexts, or to prompts about global behavior; and adapting item context to reflect more everyday scenarios. These changes are summarized in Table~\ref{tab:karaSuggestions}. The prototype that came out of the initial review is named the ``protoGERQN.''

\renewcommand{\arraystretch}{1.2}
\setlength{\tabcolsep}{4pt}
\begin{table}
    \begin{tabular}{p{0.12\textwidth} p{0.15\textwidth} p{0.15\textwidth}}
        & \textbf{PIQL} & \textbf{protoGERQN} \\
        \hline
        Construct 
            & Calc-based & Algebra-based \\
            & 20 Items & 17 Items \\
            & 45 min. & 30 min. \\
        \hline
        Covariation
            & \raggedright{Changing rates of change} & Linear rates of change \\
            & Multi-variable & Single variable \\
        \hline
        Sign Reasoning 
            & \raggedright{Scalars and Vectors} & Scalars Only \\
        \hline
        Contexts 
            & Physics \& Everyday & Everyday \\
        \hline
    \end{tabular}
    \caption{A summary comparison of the protoGERQN and the PIQL. A sample of representative items of the final GERQN and the PIQL can be found in the Appendix.}
    \label{tab:karaSuggestions}
\end{table}
\begin{figure}
    \centering
    \hfill \includegraphics[width=0.25\linewidth]{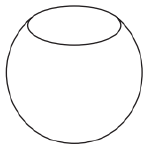} \hfill \includegraphics[width=0.25\linewidth]{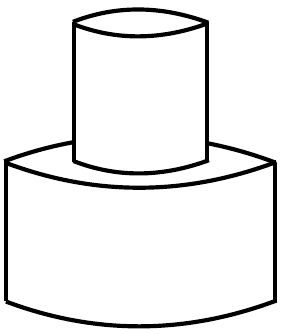} \hfill \text{} \\
    \hspace{0.25\linewidth} (A) \hfill (B) \hspace{0.25\linewidth}
    \caption{The bottle figure from (A) the PIQL and (B) the GERQN.}
    \label{fig:bottleComparison}
\end{figure}

The ``Bottle'' item is an example of an item originally in a non-linear context that we converted to a linear context (Fig.~\ref{fig:bottleComparison}). The PIQL version of this item presents the test-taker with a spherical bottle that is being filled with water. Students are asked to select the graph that correctly relates the volume of water in the bottle as a function of the height of water in the bottle. On the GERQN, the bottle has straight sides and neck.

The ``Jogger'' item is an example of an item for which we simplified the language. ``Jogger'' asks students to compare the distance traveled by two joggers; the PIQL version also asks students to identify the reasoning needed to determine which jogger went farther. For the GERQN, we removed the reasoning aspect of the responses with the intent of reducing the effort required to complete the task. In addition, we changed a numeric value in the question statement from a decimal to an integer; the middle school mathematics education expert suggested that this was more appropriate for the target population of the GERQN. This change is also aligned with research in physics education about student reasoning about integers and decimals \cite{WhiteBrahmia_Boudreaux_Kanim_2016}.

Finally, the ``Internal Energy'' and ``Money'' items on the PIQL and GERQN respectively provide an example of how we removed physics content from an item, while keeping the required reasoning the same. The PIQL item was intended to probe student understanding of symbolizing with signed quantities, using the context of the first law of thermodynamics ($\Delta U = Q+V$). It does not require understanding of the relevant quantities, meaning that work, heat, and internal energy are defined in the question stem. However, we are aware that students likely to take the GERQN include biology and chemistry majors who may be familiar with different sign conventions \cite{Hartley_Momsen_Maskiewicz_D’Avanzo_2012}. The target population may also include students for whom new physical quantities are distracting. Therefore, we created a similar item in the ``real-world'' context of money, relating the change in money in a wallet, the money spent or earned by paying for or doing a job, and the money spent or earned by selling or buying something. A full comparison of the PIQL and GERQN items described in this section can be found in the Appendix.

\begin{figure}
    \begin{framed}
    \vspace{-6em}
    \begin{minipage}[t]{0.58
    \linewidth}
    \begin{flushleft}
    \footnotesize{The graph at right represents how fast two children are growing vs time. The children are named Alex and Jordan, and their growth is measured starting on their 10th birthday when they are both the same height.}    
    \vspace{1em}
    \end{flushleft}
    \end{minipage} \begin{minipage}{0.4\linewidth}
        \vspace{6em}
        \includegraphics[width=\linewidth]{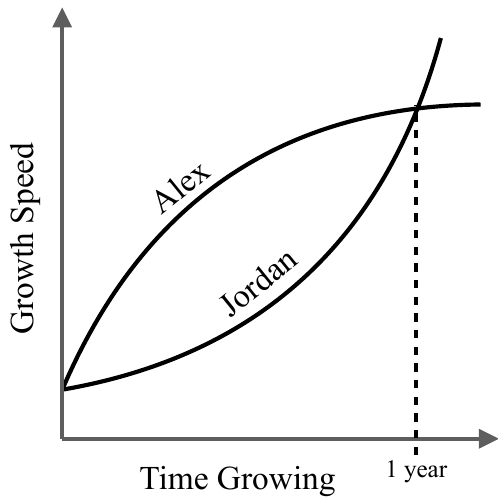}
    \end{minipage}
    \begin{minipage}{\linewidth}
    \raggedright
    \footnotesize{Which of the following choices best describes how much the children have grown in the year shown?}
        \begin{itemize}
        \footnotesize{
            \item[a.] Alex has grown more than Jordan.
            \item[b.] Jordan has grown more than Alex.
            \item[c.] Alex and Jordan have grown the same amount.
            \item[d.] The graph does not provide enough information to compare how much the two children have grown.
            }
        \end{itemize}
    \end{minipage}
    \end{framed}
    \caption{The GERQN item ``Growth,'' where students are asked to compare the growth of two children using a graph of growth rate.}
    \label{fig:plant}
\end{figure}

\subsection{Iterative Review}
Iterative review took place over two years, during which time we conducted interviews with students, conducted a faculty focus group, and administered versions of the assessment to algebra-based introductory physics courses (see Tab.~\ref{tab:dataSummary}). 

Student interviews resulted in revisions to several items, either to clarify the prompt, adjust answer options, or change the context of the problem. The final outcome of these revisions was version 1.6. The faculty focus group provided evidence that instructors consider the reasoning required by the GERQN valuable and aligned with the learning objectives of their courses.  

One unexpected outcome of the faculty focus group was a discussion of whether the ``Growth'' item (see Fig.~\ref{fig:plant}) was appropriate for an algebra-based course, or whether it required calculus to solve. After discussion, it was agreed that this item could be solved by noticing that one child grew faster than the other over the entire duration. This reasoning was confirmed in student interviews, where several students used that line of thinking to arrive at the correct answer.

Quantitative measures were used throughout the iterations to ensure that changes made did not negatively impact established statistical results from the PIQL, and as a check that all the answer options were useful to include. For example, distractors that were not chosen frequently were critically examined in student interviews, and those that were never chosen were removed. How item and inventory statistics changed across versions is described in more detail in Section~\ref{sec:valid}. A comparison of representative items from the GERQN and the PIQL, can be found in the Appendix.

\section{Validation Results}
\label{sec:valid}

In this section we describe how we validated the version of the GERQN that emerged from iterative review. Validation took place over one year, during which time we conducted individual student interviews, conducted individual interviews with experts, and administered version 3.0 in an algebra-based introductory physics sequence (see Tab.~\ref{tab:dataSummary}). Together these data form the basis from which we sought evidence of validity of the instrument and individual items, both qualitatively and quantitatively.

\begin{table}
    \centering
    \renewcommand{\arraystretch}{1.2}
    \setlength\tabcolsep{4pt}
    \begin{tabular}{|l|c|c|c|c|}
    \hline
                                     &   & Average & Average & Average \\
         \multicolumn{1}{|c|}{Course}& N & Pearson & Cronbach's & Ferguson's \\
                                     &   & Corr. Coeff. & $\alpha$ & $\delta$ \\
    \hline
    \hline
         PreMech   & 746  & 0.95 & 0.67 $\pm$ 0.05 & 0.95 $\pm$ 0.01\\
         PostMech  & 573  & 0.91 & 0.69 $\pm$ 0.06 & 0.96 $\pm$ 0.01\\
         PostEM    & 293  & 0.89 & 0.73 $\pm$ 0.07 & 0.95 $\pm$ 0.01\\
    \hline
    \end{tabular}
    \caption{Inventory statistics for v3.0. Pearson Correlations were calculated between all combinations of quarters and averaged; uncertainty represents the standard deviation across these combinations. Cronbach's $\alpha$ is calculated in aggregate across quarters; uncertainty represents width the 95\% confidence range. Ferguson's $\delta$ was calculated across all quarters.}
    \label{tab:testStats}
\end{table}

\begin{figure*}
\centering
    \includegraphics[width=0.45\textwidth]{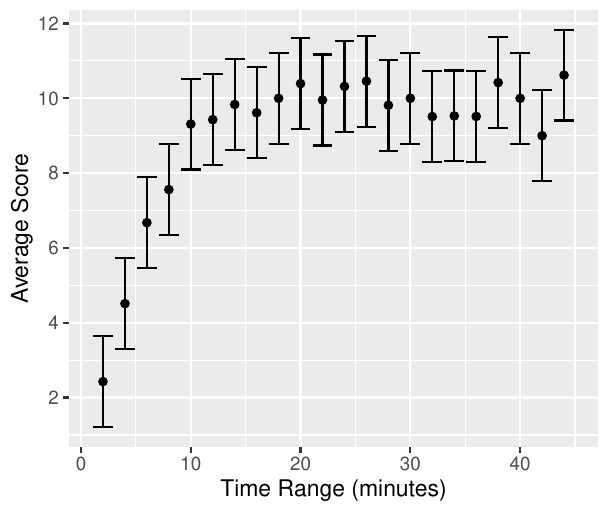}
    \includegraphics[width=0.45\textwidth]{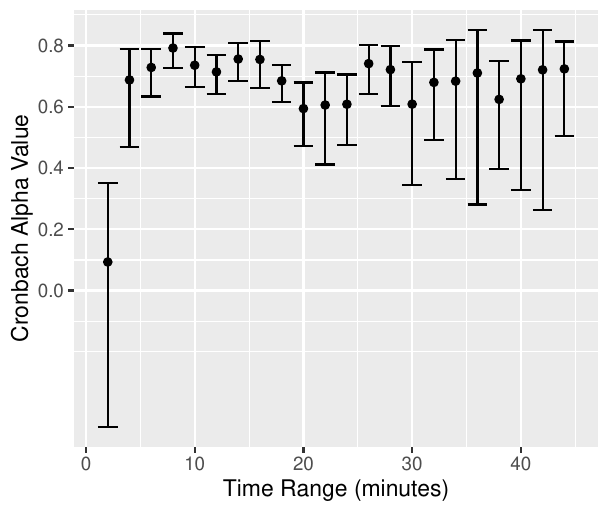} \\
    \hspace{1cm} (A) \hspace{0.4\textwidth} (B)

    \caption{(A) Average score with respect to time on task and (B) Cronbach's $\alpha$ with respect to time on task, both for GERQN version 3.0 ($N=2118$). Students who took between 0 and 45 minutes are shown\footnote{These plots show results from 1610 students who completed the test within 45 minutes. Data beyond this time are too sparse to reliably group into 2-min bins}. Error bars represent  95\% confidence intervals.}
    \label{fig:timefilter}
\end{figure*}

\subsection{Data-Driven Cut-Off Time}
Due to the out-of-class and online nature of the assessment, there was an open question of how seriously students approach the items. We binned student responses ($N = 2118$ for v3.0) into 2-minute intervals based on the time the students took on the assessment, and compared the average score and Cronbach's $\alpha$ of these groups (Fig.~\ref{fig:timefilter}). These data suggest that, for students who spend fewer than 10 minutes engaging with the test, there is a strong relationship between time on task and student score, as well as a distinct relationship between time on task and Cronbach's $\alpha$ for very low times. After about 10 minutes, we see no relationship between either of these quantities and time on task.  We use this as evidence that the data from students who spend less than 10 minutes engaging with the assessment are not representative of the overall population. The statistical analysis we present in this paper was therefore conducted with students who took at least 10 minutes on the test, and left 5 items or fewer unanswered ($N = 1612$ for v3.0, and $N = 2778$ for v1.6).\footnote{The same analysis was conducted with version 1.6, with the same result. Version 1.6 data shown in this paper are also filtered for students who took more than 10 minutes and left 5 items or fewer blank.}

\subsection{Inventory Validation}
\label{subsec:test}
During individual validation interviews, all physics instructor experts expressed that the items on the GERQN collectively represented desired learning outcomes for their introductory physics courses. They uniformly agreed that student improvement on the assessment would be a valued outcome of completing their courses, and that their students are unlikely to answer all items correctly at the very start of their courses. The emphasis of the different instructors' learning outcomes varied; some were more focused specifically on proportional reasoning than covariation more generally, for example. We interpret this spread as an indication that the test as a whole represents quantitative reasoning that is valued across different institutional contexts and different levels of physics instruction. 

The interviews also provided context for the inventory at different levels of mathematics preparation. For example, one expert teaches physics to freshman in high school in two different courses: one in which students have typically completed algebra-I and one in which students are typically co-enrolled in algebra-I. This expert expressed that the GERQN would be suitable for both, but their students who have completed algebra-I would be more comfortable with the function notation used on several items. The expert noted that students co-enrolled in algebra-I may misinterpret functions such as $N(t)$ as $N \times t$. Another expert who also teaches a physics course in which students are co-enrolled in algebra-I noted that they found the GERQN less valuable for that population than for their students who had completed algebra-I. We take these expert insights as a view into validity for the earliest physics course likely to use the instrument, and conclude that the GERQN is appropriate for students who have completed algebra-I and beyond. We suggest the GERQN be used with caution in high-school physics courses for which Algebra I is not a prerequisite. Across all experts, we confirmed that for the target population, the mathematics notation and functions used are at an appropriate level.

\begin{figure*}
    \centering
    \includegraphics[width=\linewidth]{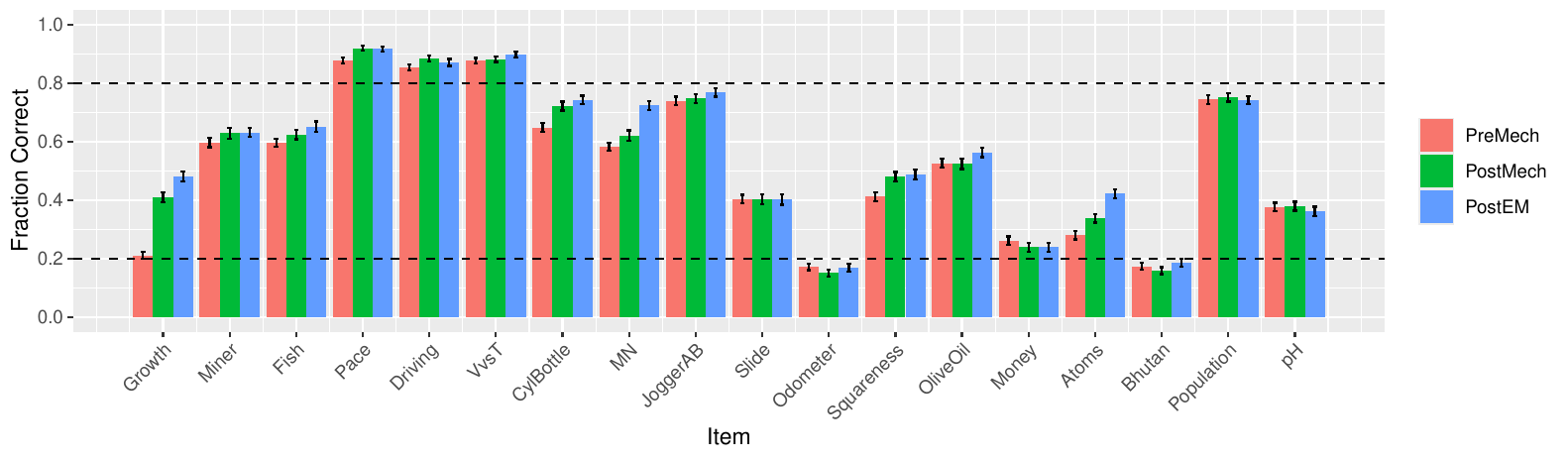} \\
    (A)
    
    \includegraphics[width=\linewidth]{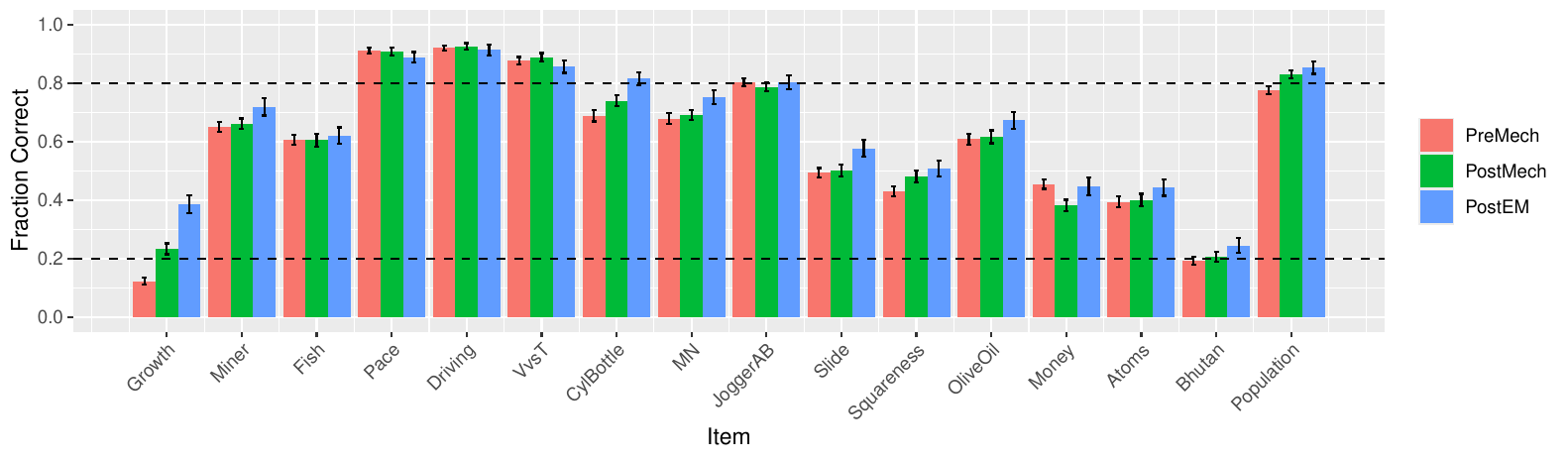} \\
    (B)
    \caption{Difficulty measures for each item on (A) version v1.6 and (B) v3.0 of the GERQN. Here, difficulty means the fraction of students who answered each item completely correctly. The horizontal lines represent the ideal range for inventory items. These data were collected over six quarters at the main institution ($N = 2,778$ for v1.6, $N = 1612$ for v3.0).}
    \label{fig:diff}
\end{figure*}

\begin{figure*}
    \centering
    \includegraphics[width=\linewidth]{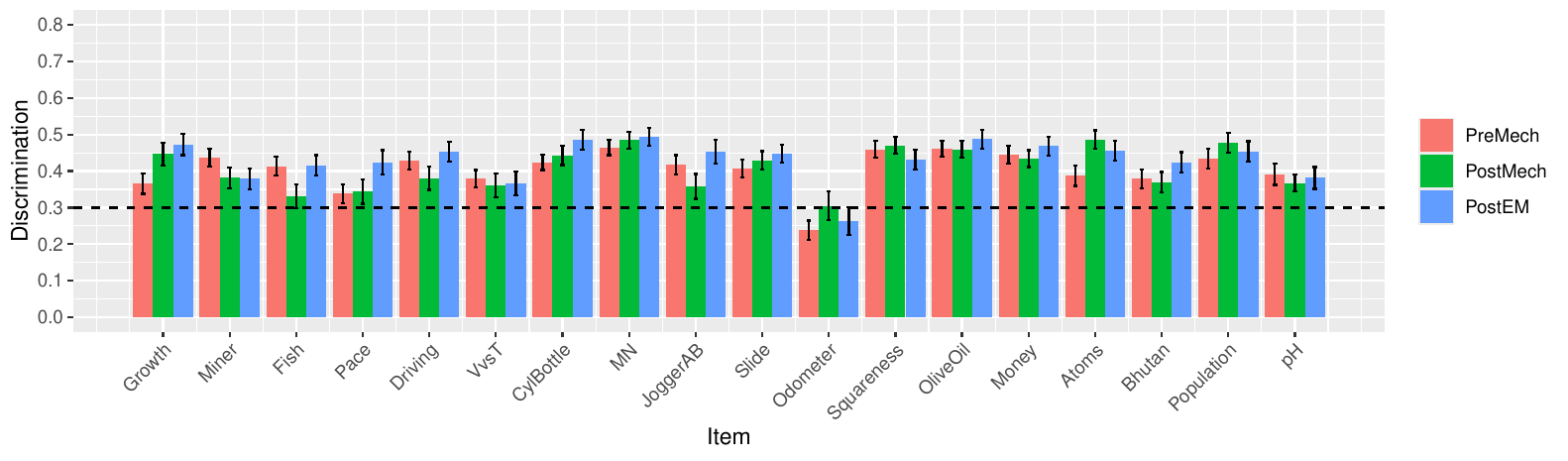} \\
    (A)
    
    \includegraphics[width=\linewidth]{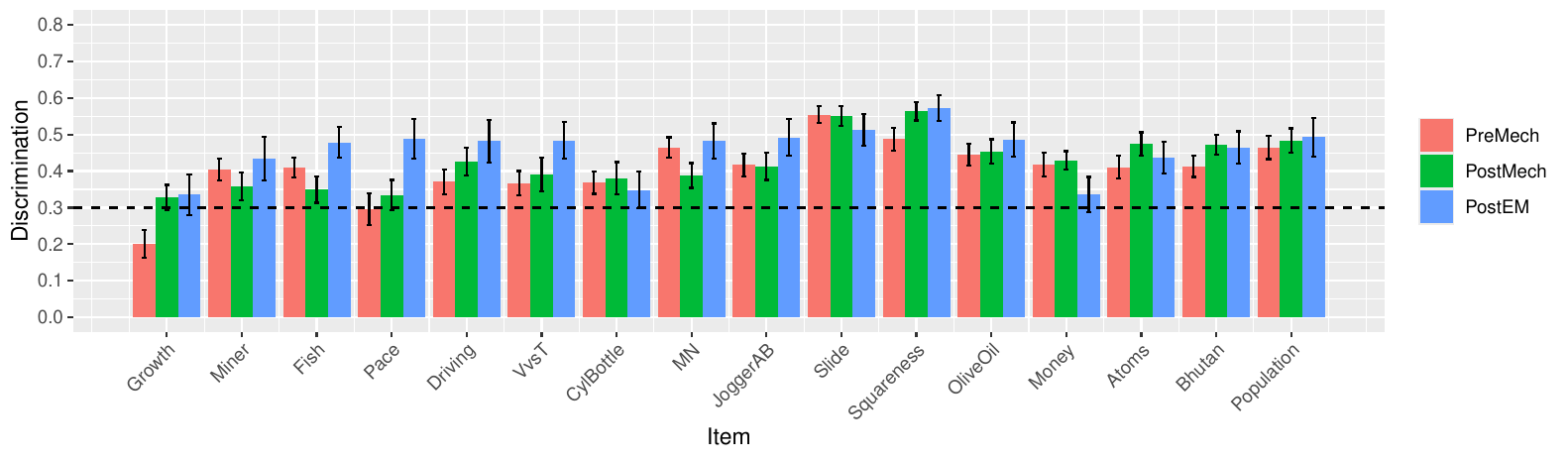} \\
    (B)
    \caption{Discrimination measures for each item on version 1.6 and 3.0 of the GERQN. The horizontal line represents the recommended minimum. These data were collected over six quarters at the main institution ($N = 2778$ for v1.6, $N = 1612$ for v3.0).}
    \label{fig:disc}
\end{figure*}

Quantitative validation of the instrument was established along the following lines:
\begin{itemize}
    \item test-retest stability via the Pearson Correlation Coefficient,
    \item internal consistency and reliability via Cronbach's $\alpha$,
    \item discriminatory power via Ferguson's $\delta$, and
    \item alignment with the single factor observed on the PIQL \cite{PIQLPaper} via exploratory factor analysis.
\end{itemize}
The results of these analyses are summarized in Table~\ref{tab:testStats}.

Test-retest stability was established through a Pearson Correlation Coefficient between two quarters of the same version, administered at the beginning of the same course. The data were aggregated by score and the percentage of students who received each score compared across quarters. The correlation indicates how similar the score frequencies are across quarters, where a 0 represents no correlation and a 1 represents a perfect correlation. Adams and Wieman report Pearson correlations are expected to be above 0.9 \cite{Adams_Wieman_2011}; we see that all courses are near or exceed this benchmark. 

Cronbach's $\alpha$ is an indicator of internal consistency. A high score indicates that item scores are related to one another. In general, a Cronbach's $\alpha$ of 0.7 or above is recommended; each course meets this benchmark within a 95\% confidence range.

Ferguson's $\delta$ can be used as a measure of test-wide discrimination. A high Fergusons's $\delta$ indicates that there is a wide spread of students across all score options. For example, Ferguson's $\delta$ is 1 if an equal proportion of students earn each possible score; it is 0 if all of the students earn the same score. It is distinct from standard deviation in that Ferguson's $\delta$ provides a measure of how students are spread across the possible scores, regardless of the mean. There is speculation as to whether Ferguson's $\delta$ is a good measure of discrimination because it is population dependent \cite{Terluin_Knol_Terwee_deVet_2009}, but we include it here as all of the data presented in this paper are reflective of the student population at the main institution. We consider validating the GERQN at other institutions an area of future work.

Prior research found that while the PIQL was built on three facets of physics quantitative literacy (reasoning about sign, proportional reasoning, and covariational reasoning), exploratory factor analysis found it to be a single-factor assessment. This finding provided evidence that PQL is a way of reasoning \cite{PIQLPaper}. We confirmed this result with GERQN v3.0 through exploratory factor analysis. We used the Kaiser-Meyer-Olkin (KMO) criterion to determine whether the data were suitable for factor analysis. We found a KMO value of 0.887, which suggests that each item is sufficiently correlated with the others such that one or more factors can be extracted \cite{Hair_Black_Babin_Anderson_2014}. Similarly, we found the Bartlett's test of sphericity was significant with an alpha value of 0.05 ($\chi^2(120) = 5697.75$, $p < .001$). We used the Kaiser-Guttman criterion and found that a maximum of four factors, and a minimum of one, could be extracted. Therefore, we modeled the data using a single factor and found a confirmatory fit index of 0.93, with RMSEA = 0.07, which aligns with recommended cut-off values \cite{Eaton_Frank_Johnson_Willoughby_2019, Hoyle_2023, Brown_2015, Bowen_Guo_2012}. These statistics suggest that the test is well described by a single factor, confirming our interpretation: for students at the introductory, algebra-based physics level, the facets of PQL are not clearly separable.

\subsection{Item Analysis}
\label{subsec:items}

We examined individual items during student and expert interviews, as well as using classical test theory, to validate that the items appear appropriate to expert instructors, are understood by students, are of the appropriate level, and have discriminatory power. 

We calculated the classical test theory difficulty and discrimination of each item for administrations of v1.6 and v3.0, seeking to meet the same standard as established on the PIQL \cite{PIQLPaper}. We aim for a wide spread in difficulties between 0.2 and 0.8, and discrimination values above 0.3. Here, the word ``difficulty'' refers to the fraction of students who choose the correct answer on a particular item; ``discrimination'' is a measure of how correlated choosing the correct answer on that item is with the overall test score (point biserial correlation). Discrimination is high when students who choose the correct answer are also likely to score high on the assessment as a whole; it is low when the correlation is not as strong. The results for v1.6 and 3.0 across the introductory sequence are shown in Figures~\ref{fig:diff} and~\ref{fig:disc}. We used these quantitative measures for v1.6 and v3.0 to inform final decisions about items to keep and items to remove. 

Student interviews were conducted to establish evidence that:
\begin{itemize}
    \item the students interpreted the questions and answer choices as intended, and that
    \item there were no commonly desired answers that were not already multiple choice options.
\end{itemize}
We also sought to confirm prior characterization of student reasoning associated with incorrect answer choices, toward helping instructors interpret the ways that their students might be reasoning. Across all interviews, students understood the items as intended and chose answers for the reasons we expected. Only one item was revised as a result of these interviews; the change was made half-way through the interviews, and confirmed with the second half. 

Several experts had item-specific suggestions to improve readability. They also provided feedback on which items would be too difficult or confusing for their students, so that they did not expect valuable data on those items if the inventory were to be administered in their classes. 
This feedback led to some revisions of some items, and the removal of items that were considered too difficult. For example, in version 2.0, the ``Growth'' item curves both intersected the origin (Fig.~\ref{fig:plant}). The mathematics education researcher whom we interviewed noted that it would be a more valuable item if the curves had a non-zero intercept. The change in difficulty for this item between versions (see Fig.~\ref{fig:diff}) suggests that students may have been answering this item correctly for the wrong reasons before we made the change. Other items were revised for language to clarify assumptions of the simple models presented; experts suggested that students may find simplicity unnecessarily distracting. Their suggestions were confirmed with student interviews. Finally, most of the experts we interviewed agreed that one item (``Odometer'') was too hard, and it was removed; this is supported by student data, with fewer than 20\% of students answering ``Odometer'' correctly on v1.6 (Fig.~\ref{fig:diff}), and ``Odometer'' being the only item to consistently have a discrimination value below 0.3 (Fig.~\ref{fig:disc}). Another item (``pH'') was removed because it did not represent essential reasoning.

\begin{figure}
    \begin{framed}
    \vspace{-6em}
    \begin{minipage}[t]{0.58
    \linewidth}
    \begin{flushleft}
    \footnotesize{A person has a flag of Bhutan, shown at right. Another person has a larger flag of Bhutan. The larger flag is 2 times wider and 2 times taller than the smaller flag.}    
    \vspace{1em}
    \end{flushleft}
    \end{minipage} 
    \begin{minipage}{0.4\linewidth}
        \vspace{6em}
        \includegraphics[width=\linewidth]{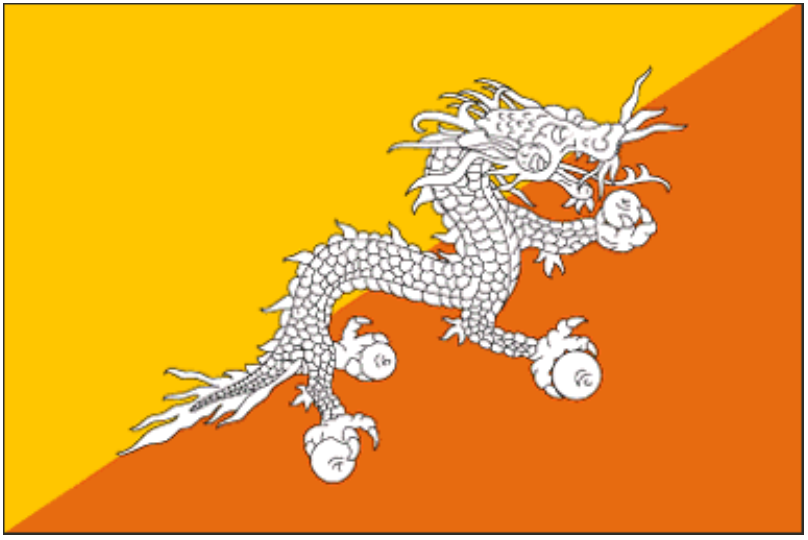}
    \end{minipage}
    \begin{minipage}{\linewidth}
    \RaggedRight
    \footnotesize{Which of the quantities below are 2 times larger for the larger flag compared to the smaller flag? \textbf{\textit{Choose all that apply.}}}
        \begin{itemize}
        \footnotesize{
            \item[a.] The distance around the edge of the flag.
            \item[b.] The amount of cloth needed to make the flag.
            \item[c.] The length of the curve forming the dragon's backbone.
            \item[d.] The diagonal of the flag.
            \item[e.] None of these quantities.
            }
        \end{itemize}
    \end{minipage}
    \end{framed}
    (A) \\
    
    \includegraphics[width=0.45\textwidth]{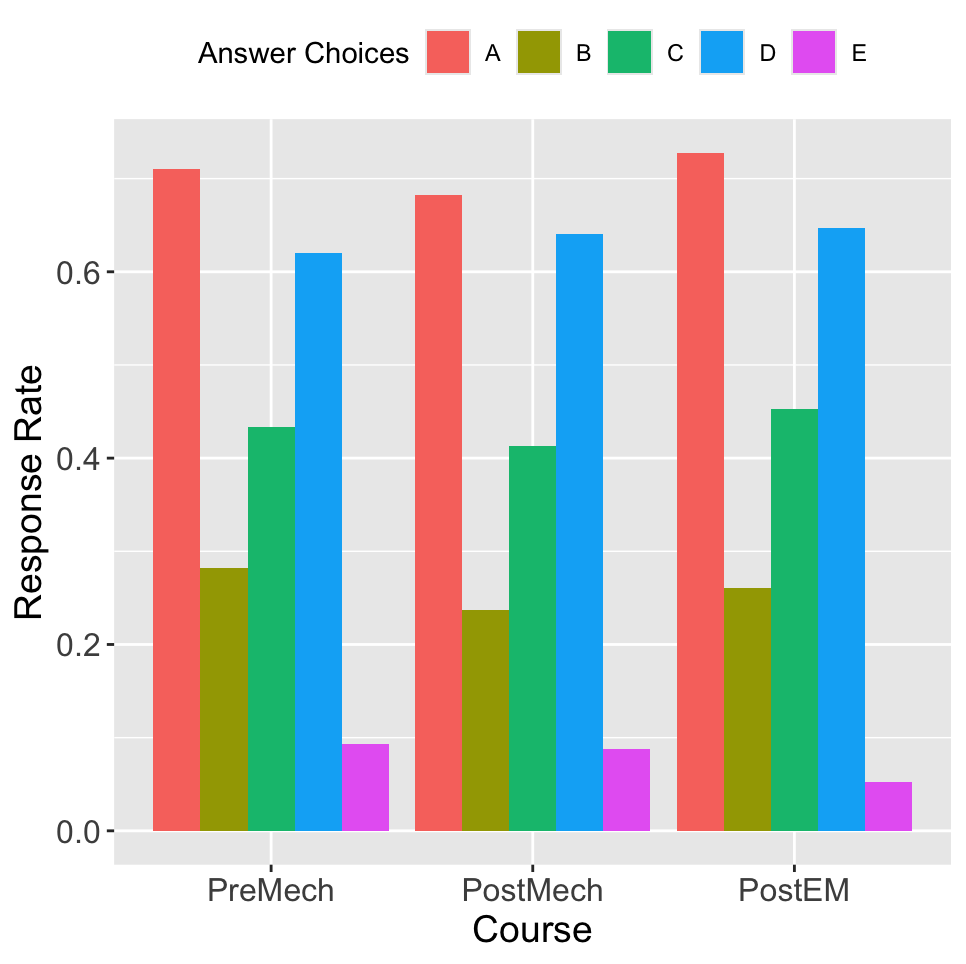} \\
    (B) 
    \caption{(A) The Bhutan item, and (B) associated rates of student responses to each answer choice, treated independently. Bhutan did not change between versions; this plot includes data from both v1.6 and 3.0 (N $=4231$).}
    \label{fig:bhutanResponses}
\end{figure}

A subset of experts also noted that a second item, ``Bhutan,'' would likely be too hard for their students (Fig.~\ref{fig:bhutanResponses}A). This was confirmed with test statistics (Fig.~\ref{fig:diff}). However, Bhutan is a multiple-choice-multiple-response (MCMR) item. Figure~\ref{fig:diff} only shows the fraction of completely correct responses for MCMR items; Figure~\ref{fig:bhutanResponses}(B) suggests that many more students are choosing a subset of correct answers. In addition, among the experts who had concerns about Bhutan, the concern was that answer option (C) was too challenging. All the experts agreed it would be valuable to them if their students selected more correct answer choices and fewer incorrect answer choices, even if the students did not get the item completely correct. They also agreed that students' ability to reason about scaling in quantities that are not well described by a familiar equation (something that is measured by answer choice C) is a valued learning outcome of introductory physics for the target population. Therefore, we decided to keep this item in the inventory.

We chose to keep several items that have a higher than recommended difficulty ($>80\%$ of students choosing the correct answer). The students at the main institution typically have more access to prior mathematics instruction than the target population for the GERQN as a whole. We consider these items to be a positive feature of the test, and we expect the item statistics to be different for data collected in other educational settings. 

\section{Implications for Instruction}
\label{sec:conclusion}

One key goal in developing and validating the GERQN was to provide instructors with a practical tool to support PQL instruction. The GERQN can help identify meaningful learning objectives, track shifts in student performance over time, and highlight changes in specific skills through item-by-item analysis. Instructors can use it to better understand how their students reason by:
\begin{enumerate}
    \item Monitoring average scores across cohorts or instructional periods
    \item Analyzing individual items to identify specific areas of difficulty
\end{enumerate}

\begin{figure}
    \centering
    \includegraphics[width=0.85\linewidth]{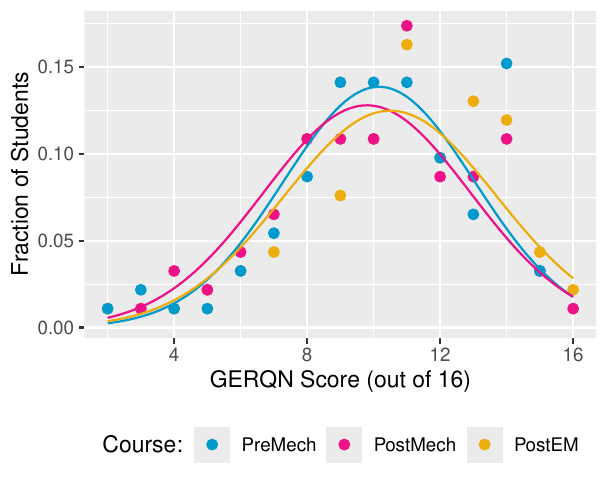}
    \caption{GERQN score distribution over the year-long introductory algebra course at the main institution. Only students who spent longer than 10 minutes and completed the test in each course are shown; these data are from v3.0. Data are fitted with a normal distribution. ($N = 92$)}
    \label{fig:gerqnTrends}
\end{figure}

Initial findings at the main institution echo results from the PIQL: PQL does not significantly improve through traditional physics instruction alone (Fig.~\ref{fig:gerqnTrends}). The GERQN thus offers a valuable tool to help instructors address these persistent challenges within their own classrooms. 

Aligning early PQL development with students’ recent math coursework can boost student confidence and sense of agency in applying mathematical reasoning in future courses. This is especially important for students from lower socioeconomic status (SES) districts, where access to physics and calculus in high school is often limited \cite{WhiteBrahmia_Cochran}. Algebra-based physics is especially common as a first physics course in high school. Pre-college teachers---experienced in designing creative, student-centered learning---could use the GERQN to develop engaging approaches to PQL instruction. 

During validation, several early-career college faculty expressed strong interest in supporting the reasoning assessed by the GERQN but also highlighted the need for professional development to deepen their own understanding of PQL. Precollege teachers would also benefit from this type of dedicated professional development. The GERQN has the potential to serve as a foundation for such efforts. Initially validated with algebra-based physics students at an R1 institution, it is now being tested across a range of post-secondary contexts, including two-year colleges, minority-serving institutions, and rural campuses.

Future efforts will focus on validating the GERQN in pre-college classrooms, creating professional development resources, and designing replicable workshops for both college and K–12 educators. Of course, no single assessment can fully address the longstanding challenges students face in developing PQL. We view the GERQN as a catalyst for informed instruction that can help move the needle on this way of reasoning. 

\section{Acknowledgments}
The authors thank Andrew Boudreaux and Stephen Kanim for their intellectual contributions to this work developing the framework and assessment items, Michael Loverude for contributions to our investigations of covariational reasoning, and those that participated in interviews for their thoughtful feedback. We would also like to thank our project advisory board members: Mehri Fadavi, Cameron Byerly, and Elizabeth Schoene. We are grateful to the contributions and support of the University of Washington Physics Education Group and the leadership of the introductory physics courses --  David Smith, Nikolai Tolich, Kazumi Tolich, and Peter Shaffer in particular for making data collection possible over several years. This work was supported by the National Science Foundation under grants DUE-2214283 and DGE-1762114.
\vfill

\pagebreak
\onecolumngrid
\section{Appendix}
Here we provide a comparison of a sample of items in the PIQL (available on PhysPort) and GERQN v3.0 for the interested reader. 

\renewcommand{\arraystretch}{1.2}
\setlength{\tabcolsep}{6pt}
\begin{longtable}{p{0.2\textwidth}p{0.35\textwidth}p{0.35\textwidth}}
\hline\hline\hline
    \textbf{Item and Summary} 
    & \textbf{PIQL \cite{PIQLPaper}} 
    & \textbf{GERQN v3.0} \\
\hline
\endfirsthead
\hline\hline\hline
    \textbf{Item and Summary} 
    & \textbf{PIQL \cite{PIQLPaper}} 
    & \textbf{GERQN v3.0} \\
\hline
\endhead
\hline\hline
\multicolumn{3}{c}{\textit{Con't next page\ldots}}\\
\endfoot
\hline\hline
\endlastfoot
\multicolumn{3}{l}{\textbf{Growth}} \\
    Context was adjusted, and ``rate'' was changed to ``speed''. The functions were also moved off the origin.
    & The graph shown represents the \emph{growth rate} vs.\ \emph{time} for two plants. Which of the following statements best describes the growth of the two plants from $t=0$ to $t=1$~month? 
    & The graph at right represents how fast two children are growing vs time. The children are named Alex and Jordan, and their growth is measured starting on their 10th birthday. Which of the following choices best describes how much the children have grown in one year?\\

    & \includegraphics[height=1.5in]{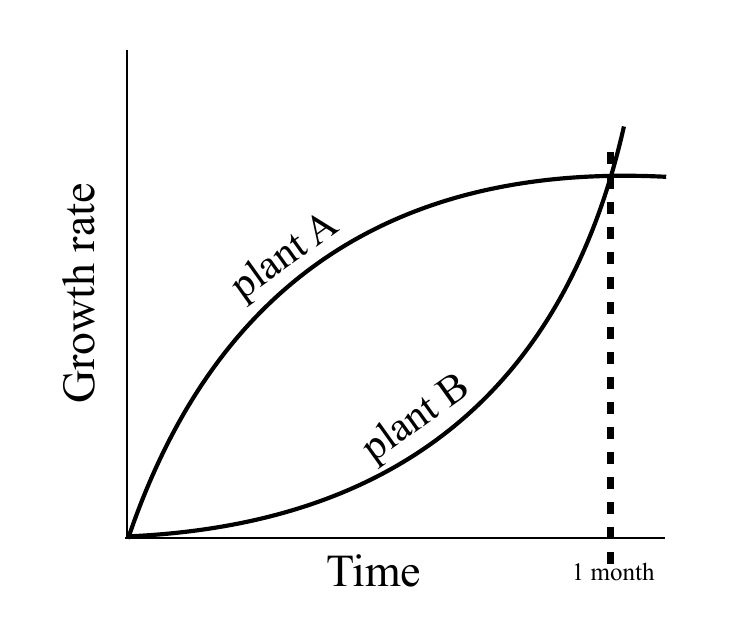}
    & \includegraphics[height=1.5in]{figures/childrenGrowing.pdf} \\

    & \begin{itemize}
        \item[a.] Plants A and B have the same amount of growth.
        \item[b.] Plant A has experienced more growth than plant B.
        \item[c.] Plant B has experienced more growth than plant A.
        \item[d.] The graph does not provide enough information to compare the growth of the two plants.
    \end{itemize}
    & \begin{itemize}
        \item[a.] Alex and Jordan have grown the same amount.
        \item[b.] Alex has grown more than Jordan. \newline
        \item[c.] Jordan has grown more than Alex. \newline
        \item[d.] The graph does not provide enough information to compare how much the two children have grown.
    \end{itemize} \\

\hline
\multicolumn{3}{l}{\textbf{Bottle}} \\
    The spherical bottle was changed to a cylindrical one with two sections of different diameter, effectively linearizing the problem.
    & Assume that water is poured into a spherical bottle at a constant rate. Which of the following graphs best represents the height of the water, $h$, in the spherical bottle as a function of the amount of water in the bottle, $V$?
    & Water is poured into an empty bottle until it is full. The bottle is shaped like two cylinders, as shown at right. Which graph best represents the height of the water in the bottle, $h$, as a function of the amount of water in the bottle, $V$? \\

    & \includegraphics[height=0.75in]{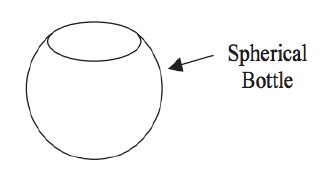}
    & \includegraphics[height=0.75in]{figures/cylBottleNoWords.pdf} \\

    & \text{} \newline \begin{enumerate*}
        \item[a.] \includegraphics[width=0.5in]{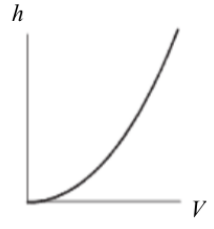}
        \item[b.] \includegraphics[width=0.5in]{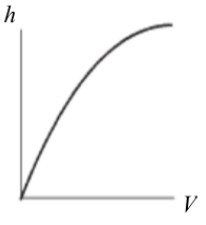}
        \item[c.] \includegraphics[width=0.5in]{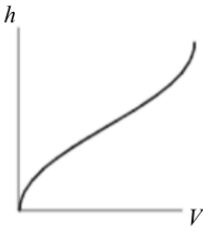} \newline \newline
        \item[d.] \includegraphics[width=0.5in]{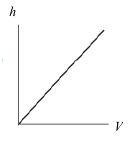}
        \item[e.] \includegraphics[width=0.5in]{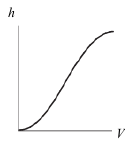}
    \end{enumerate*} 
    & \text{} \newline \begin{enumerate*}
        \item[a.] \includegraphics[width=0.5in]{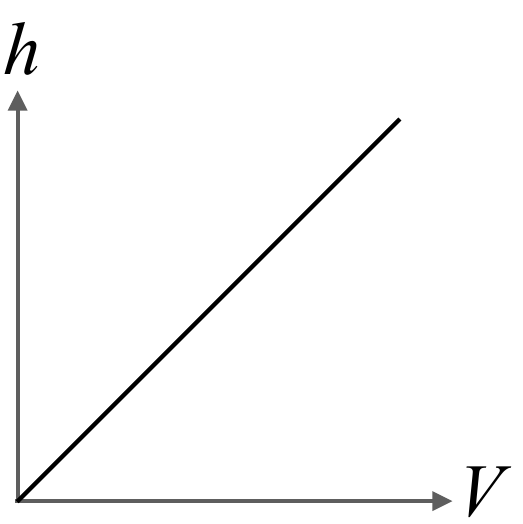}
        \item[b.] \includegraphics[width=0.5in]{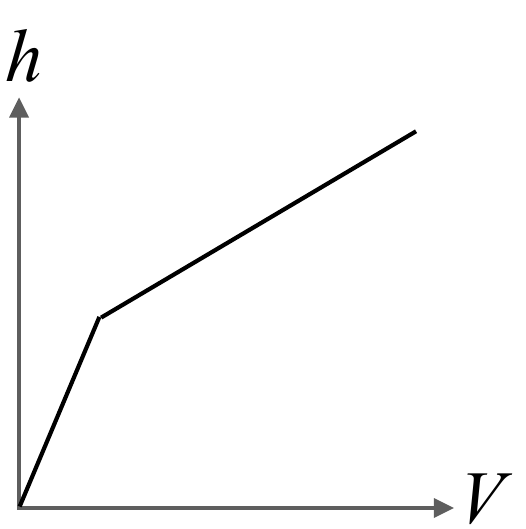}
        \item[c.] \includegraphics[width=0.5in]{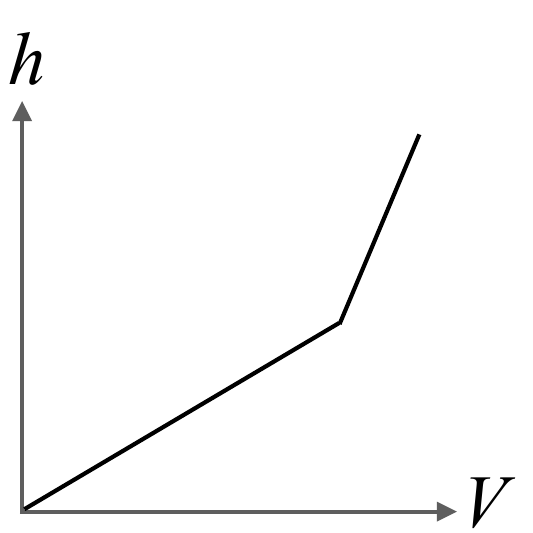} \newline \newline
        \item[d.] \includegraphics[width=0.5in]{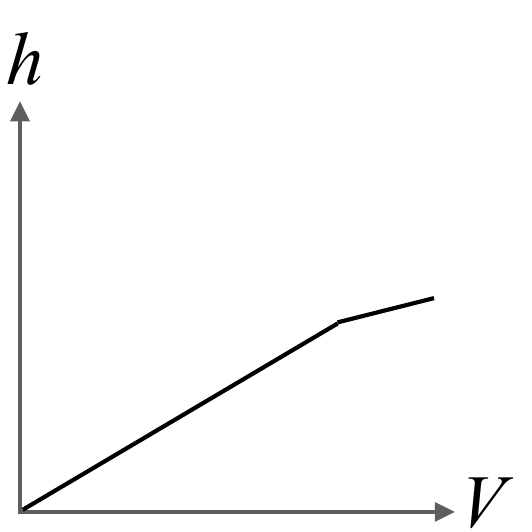}
        \item[e.] \includegraphics[width=0.5in]{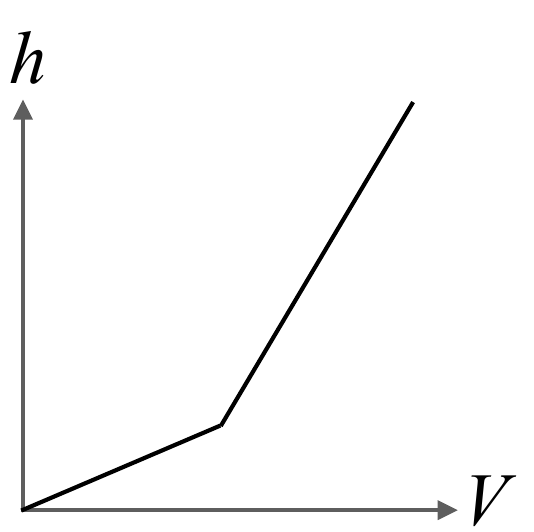}
    \end{enumerate*} \\

    \pagebreak

\multicolumn{3}{l}{\textbf{Fish}} \\
    Function adjusted to be at the appropriate mathematical level. Question stem includes more description to account for the discrete nature of the new function.
    & The wildlife game commission released 500 fish into a lake. The function $N(t)$ defined by \newline $N(t)=\frac{600t+500}{0.5t+1}$ \newline represents the approximate number of fish in the lake as a function of time (in years). Which one of the following best describes how the number of fish in the lake changes over time?
    & 500 fish are released into a lake. The number of fish in the lake are recorded at the end of each year. In the first year, the number of fish gets smaller. After the first year, $N(t)$ represents the number of fish recorded, and $t$ represents the number of years since the 500 fish were originally released into the lake. The equation for $N(t)$ is \newline $N(t)= 500-\frac{30}{t}$, starting at $t = 1$. \newline Which of the following choices best describes the trend in the fish population after the first year, over a period of many, many years? \\

    & \begin{itemize}
	\item[a.] The number of fish gets larger each year, but does not exceed 500.
	\item[b.] The number of fish gets larger each year, but does not exceed 1200.
	\item[c.] The number of fish gets smaller each year, but does not get smaller than 500.
	\item[d.] The number of fish gets larger each year, but does not exceed 600.
	\item[e.] The number of fish gets smaller each year, but does not get smaller than 1200.
    \end{itemize}
    & \begin{itemize}
	\item[a.] The number of fish keeps getting smaller until the fish are gone.
	\item[b.] The number of fish keeps getting smaller but does not drop below 470.
	\item[c.] The number of fish eventually grows to nearly 500 again.
	\item[d.] The number of fish eventually grows to a number greater than 500.
    \end{itemize} \\

\hline
\multicolumn{3}{l}{\textbf{Inverse G / Pace}} \\
    Replaced a changing rate of change with a constant rate of change.
    & Near the surface of Earth, the acceleration due to gravity is 9.8 m/s$^2$ (note: $\frac{\textrm{m}}{\textrm{s}^2} = \frac{\textrm{m/s}}{\textrm{s}}$). For objects in vertical free fall near the surface of Earth, the number 9.8 provides the following information: \textit{the speed of the object will change by 9.8 m/s during each second of its motion.} Consider the reciprocal of this number, 0.1 s$^2$/m  (note: $\frac{\textrm{s}^2}{\textrm{m}} =\frac{\textrm{s}}{\textrm{m/s}}$). For objects in vertical free fall near the surface of Earth, what specific information does this quantity (0.1 s$^2$/m) convey? Select the single best choice below.
    & A person runs at 3 m/s, meaning that the runner moves 3 meters each second they are running. The reciprocal of 3 m/s is 0.33 s/m. Which of the following choices best describes the specific information that 0.33 s/m tells us about the motion of the person? \\

    & \begin{itemize}
	\item[a.] The speed of the object will change by 0.1 m/s during each second.
	\item[b.] The motion of the object will change by 0.1 s$^2$ in each meter.
	\item[c.] It takes 0.1 s for each 1 m/s change in the object's speed.
	\item[d.] It takes 0.1 s$^2$ for the object to fall each meter.
	\item[e.] None of these make sense; the number 0.1 does not have a valid interpretation in this context.
    \end{itemize}
    & \begin{itemize}
	\item[a.] It takes the runner 0.33 seconds to move 1 meter.
	\item[b.] It takes the runner 1 second to move 0.33 meters.
	\item[c.] It takes the runner 0.33 seconds to move 0.33 meters.
	\item[d.] The runner's speed is 0.33 s/m. \newline
        \item[e.] None of these make sense; the quantity 0.33 s/m does not have a specific meaning.
    \end{itemize} \\

    \pagebreak

\multicolumn{3}{l}{\textbf{Jogger}} \\
    Numerals and language simplified; reasoning statements removed from the answers.
    & Joggers A and B start running at the same time from the same location. Jogger A is slower than jogger B ($0.6$ times the speed of jogger B) but runs for a longer time ($1.5$ times the amount of time that jogger B runs). How does the distance traveled by A compare to the distance traveled by B? Select the answer with the best reasoning.
    & Joggers A and B start running at the same time from the same location. Jogger A is slower than jogger B ($0.6$ times the speed of jogger B) but runs for twice as much time as jogger B.  How does the distance traveled by A compare to the distance traveled by B? \\

    & \begin{itemize}
        \item[a.] The distance traveled by A is greater than B because A runs for more time.
        \item[b.] The distance traveled by B is greater than A because B runs faster.
        \item[c.] They both run the same distance because although A runs for more time, B runs faster and it balances out.
        \item[d.] The distance traveled by A is greater because although B runs faster, A runs long enough that he passes B and keeps going once B has stopped.
        \item[e.] The distance traveled by B is greater because although A runs for more time, A doesn't run long enough to travel as much distance as B traveled before she stopped.
    \end{itemize}
    & \begin{itemize}
        \item[a.] The distance traveled by A is greater than B.
        \item[b.] The distance traveled by B is greater than A.
        \item[c.] They both run the same distance.
        \item[d.] There's no way to tell without knowing their speeds.
    \end{itemize} \\

\hline
\multicolumn{3}{l}{\textbf{Internal Energy / Money}} \\
    Context changed to money, variables are more explicitly defined, and fewer answer options are provided.
    & The internal energy of a system can be increased by doing positive work on the system or by heating it, and it can be decreased by cooling the system or if the system does work. Which of the following equations represent(s) this relationship ($U$ is the internal energy of the system, $Q$ is positive when energy flows into the system, and $W$ is positive when positive work is done on the system)? \textbf{\textit{Choose all that apply.}}
    \newline \newline
    \begin{itemize}
        \item[a.] $\Delta U = Q - W$
        \item[b.] $\Delta U = - Q + W$
        \item[c.] $\Delta U = Q + W$
        \item[d.] $-\Delta U = Q + W$
        \item[e.] $-\Delta U = Q - W$
        \item[f.] $-\Delta U = - Q + W$
    \end{itemize}
    & $\Delta M$ represents the change in the amount of money that is in your wallet.
        \begin{itemize}
            \item The value of $\Delta M$ is greater than zero when you receive money. 
            \item The value of $\Delta M$ is less than zero when you spend money.
        \end{itemize}
        $J$ represents the money paid for doing a job: 
        \begin{itemize}
            \item The value of  $J$ is greater than zero if you are paid to do a job.
            \item The value of  $J$ is less than zero if you pay for someone else to do a job.
        \end{itemize}
        $G$ represents the money traded for goods:
        \begin{itemize}
            \item The value of $G$ is greater than zero if you sell something.
            \item The value of $G$ is less than zero  when you buy something.
        \end{itemize}
        Which of the following equations fully represent(s) the relationship between $\Delta M$, $J$, and $G$ if there is one job done and one trade made?  \textit{\textbf{Choose all that apply.}}
        \begin{itemize}
            \item[a.] $\Delta M = J + G$
            \item[b]. $\Delta M = J - G$
            \item[c.] $\Delta M = -J + G$
            \item[d.] $\Delta M = - J  - G$
        \end{itemize}  \\

\end{longtable}
\twocolumngrid

\bibliography{references}
\end{document}